\documentclass[12pt]{article}
\usepackage{graphicx}
\usepackage{cite}
\newcommand{\be}{\begin{equation}}
\newcommand{\ee}{\end{equation}}
\newcommand{\beq}{\begin{eqnarray}}
\newcommand{\eeq}{\end{eqnarray}}
\newcommand{\ba}{\begin{array}}
\newcommand{\ea}{\end{array}}
\newcommand{\crs}{cross section}
\newcommand{\crss}{\crs s}

\newcommand{\gpa}{\gamma p\to a_0(980)p}
\newcommand{\gpf}{\gamma p\to f_0(980)p}

\newcommand{\rpt}{\rule{0pt}{14pt}}
\newcommand{\rptt}{\rule{0pt}{16pt}}

\newcommand{\ega}{E_{\gamma}}
\newcommand{\svg}{sv\gamma}
\newcommand{\ssvg}{SV\gamma}
\newcommand{\eps}{\epsilon}

\newcommand{\half}{\frac{1}{2}}
\newcommand{\bfaf}{\mbox {\boldmath $a_0\!-\!f_0$}}

\newcommand{\gamu}{\gamma^{\mu}}

\newcommand{\bfk}{\mbox {\boldmath $k$}}
\newcommand{\beps}{\mbox {\boldmath $\eps$}}

\date{}

\begin{document}
\title{\normalsize\bf Photoproduction and mixing effects of scalar
	$a_0$ and $f_0$ mesons}

\author{\normalsize  
V.~E.~Tarasov$^1$, W.~J.~Briscoe$^2$, W.~Gradl$^{\,3}$,
	A.~E.~Kudryavtsev$^{1,2}$, I.~I.~Strakovsky$^2$}

\maketitle

\vspace{-9mm}

\centerline{\small $^1$\it {Institute of Theoretical and Experimental Physics, Moscow, Russia}}
\centerline{\small $^2$\it {The George Washington University Institute for Nuclear Studies, Washington, DC 20052, USA}}
\centerline{\small $^3$\it {Johannes Gutenberg-Universit\"at Mainz, Institut f\"ur Kernphysik, D-55099 Mainz, Germany}}

\vspace{-4mm}
\begin{abstract}
The photoproduction processes $\gpa$ and $\gpf$~ at energies close
to threshold are considered. These reactions are studied in the
$\pi\pi p$, $\pi\eta p$, and $K\bar K p$ channels. Production cross
sections are estimated in different models. The role of the
$a^0_0-f_0$ mixing is examined in the invariant $\pi\pi$-,
$\pi\eta$-, and $K\bar K$-mass spectra.
\end{abstract}



\vspace{5mm}
\centerline{\bf 1.~Introduction}
\vspace{3mm}

The light scalar mesons $a_0(980)$ and $f_0(980)$ have long been
of special interest, since their nature has not, until recently,
been well understood. Their description as $q\bar q$ states in
quark models encounters difficulties since these predict the
lowest $^3P_0$ states above 1 GeV, see, e.g., Ref.~\cite{Torn}.
On the other hand, the four-quark states $q^2\bar q^2$ around
1~GeV are expected to be possible~\cite{Torn}, due to the strong
attraction between diquark and antidiquark. The four-quark
structure of scalar mesons was widely considered as compact
$q^2\bar q^2$ states~\cite{Jaffe,Achas}, or as hadronic
molecular $K\bar K$ states~\cite{Weinst,Close,Achas1}. The
latter version is inspired by the proximity of the $a_0(980)$
and $f_0(980)$ states to the $K\bar K$ thresholds together with
the established strong couplings to the $K\bar K$ channel.
In Ref.~\cite{Baru}, a model-independent approach based on the
work of Weinberg (see Ref.~\cite{Torn}) was developed for the
case of these scalars. This has led to the conclusion that they
are not pure elementary particles, but have a sizable admixture
of a molecular $K\bar K$ state, which dominates in the $f_0(980)$
case. 

In Ref.~\cite{AchaKK}, the radiative decay $\phi(1020)\to\gamma
a_0/f_0$ was suggested as a tool to reveal the nature of the 
scalars $a_0$ and $f_0$. The experimental data ~\cite{SND} on 
these decays point to a sizable $K\bar K$ component in these 
states. The $\phi(1020)$ decays $\phi\to\gamma S$ ($S=a_0,f_0$) 
and the two-photon decays $S\to\gamma\gamma$ were also considered 
in Refs.~\cite{KKNHH} and \cite{HKKN}, respectively, assuming the
molecular $K\bar K$ structure of $a_0$ and $f_0$. The decay
rates for $\phi\to \gamma S$ and $S\to\gamma\gamma$ were found
in agreement with existing data, i.e., the molecular picture
has been successfully tested for these processes.  The decay
rates of transitions $S\to\gamma\rho/\omega$ were estimated in
Ref.~\cite{KKNHH1}, and were found to be very sensitive to the
model assumed for the scalars (quark compact states or $K\bar K$
molecules).

There is also an interesting question concerning the mixing of
the isovector $a_0(980)$ and isoscalar $f_0(980)$. The known
hadronic decays of these mesons are $a_0(I=1)\to \pi\eta, K\bar K$
and $f_0(I=0)\to \pi\pi, K\bar K$.  The isospin-breaking (IB)
$a_0-f_0$ mixing (for neutral $a^0_0$), going through the common
$K\bar K$ decay channel, was suggested long ago in Ref~\cite{Achas79};
the effect occurs owing to the mass difference of neutral and charged
kaons.  This mechanism should dominate in the case of molecular
structure of scalars. Thus, the $a_0-f_0$-transition amplitude,
extracted from the experiments, also will help us to establish the
nature of these scalars.

The $a_0-f_0$-mixing effect was discussed in different processes,
i.e., $\gamma p\to\pi\pi p,K\bar Kp$~\cite{Kerb},
$\pi^-p\to\pi^0\eta n$~\cite{Achas2}, $pp\to p(\pi^0\eta)p$ (central
region)~\cite{Kirk}, $pn\to d\pi^0\eta$~\cite{KudTar,KTHHS},
$dd\to(\pi^0\eta)^{4\!}He$~\cite{Kond} (and Ref.~\cite{KudTar},
arXiv version), and $J/\Psi\to\phi\pi^0\eta$~\cite{Hanhart,Wu,Roca}.
The last two processes, forbidden in the isospin-conserving limit, are
proportional to the mixing amplitude squared, while the others are 
sensitive to the $a_0-f_0$ mixing through some differential 
observables.  First experimental results in the $J/\Psi\to\phi\pi^0\eta$ 
channel were obtained by the BES-III collaboration~\cite{Ablikim:2011}.  
Recently this collaboration has also observed the isospin-violating decay
$\eta(1405)\to\pi^0 f_0(980)$~\cite{Ablikim:2012}. This process, also
related to the charged-neutral kaon mass difference and $a_0-f_0$ mixing,
was theoretically studied in Refs.~\cite{Wu1,Aceti}.

Note that in the case of $\gamma$-induced processes, it looks difficult
to identify isospin-violating final states, since the initial photon
can be treated as isospin-0 as well as isospin-1 particle.
Thus, in the case of photoproduction processes $\gamma p\to (a_0/f_0)p\,$
considered in the present paper, it is more promising to study the IB
effects, which comes from the sharp mass behavior of the
$a_0-f_0$-transition amplitude predicted by the $K\bar K$ mechanism.

In the present paper, we consider the $a_0(980)$ and
$f_0(980)$-photoproduction processes at photon-beam energies of
$\ega\sim 1.6$~GeV. This value is quite close to the maximal energy
available at the MAMI-C facility, and is enough to produce the meson
system with an effective mass somewhat above the $K\bar K$ thresholds
to study the mixing effect discussed. This is the region of threshold
production of both $a_0$ and $f_0$ mesons  with their nominal masses.
Our consideration has much in common with that given in Ref.~\cite{Kerb},
but includes estimations of absolute cross sections and uses improved
$a_0/f_0$ parameters.

\vspace{2mm}
The paper is organized as follows. In Section~2, we describe the
resonance amplitudes for the reactions $\gamma p\to \pi\pi p,\pi\eta
p,K\bar K p$, arising from the $a_0$- and $f_0$-production amplitudes.
In Section~3, we perform the results of our calculations.  In
Subsection~3.1, we give the predictions for the total cross sections
of the processes mentioned above in different models. In Subsection~3.2,
we present the results for two-meson ($\pi\pi$, $\pi\eta$, $K\bar K$)
effective mass spectra with special attention to the $a^0_0-f_0$
mixing effect. Section~4 is the Conclusion.

\vspace{5mm}
\centerline{\bf 2.~Amplitudes}
\vspace{2mm}

\begin{figure}
\begin{center}
\includegraphics[width=3.7cm, keepaspectratio]{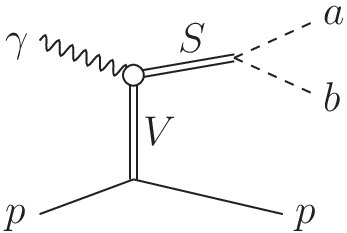} 
\end{center}

\vspace{-4mm}
\caption{Vector-meson-exchange (VME) diagrams for the reaction
         $\gamma p\to Sp\to (ab)p$.
  Wavy, solid and dashed lines correspond to the photon, nucleons
  and final $a$ and $b$ mesons, respectively. Double lines
  correspond to scalar ($S$) and vector ($V$) mesons.}\label{fig:1}
\end{figure}

\begin{figure}
\begin{center}
\includegraphics[width=6.0cm, keepaspectratio]{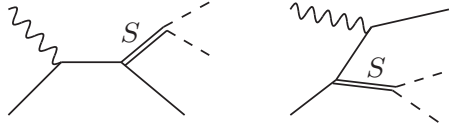}~~~~~~~~
\end{center}

\vspace{-4mm}
\caption{Born diagrams for the photoproduction reaction
         $\gamma p\to Sp\to (ab)p$ of neutral scalars $S=a^0_0,\,f_0$.
         See the notations in Fig.~\protect\ref{fig:1}.}
	\label{fig:2}
\end{figure}

Different models for $a_0$- and $f_0$-meson production can be
considered. One is that derived by Oset and coauthors~\cite{Marco}
(see also Refs.~\cite{Roca,Aceti} and references therein) and
based on the chiral unitary approach.
Another model was proposed
in Ref.~\cite{Donn}, in which scalar mesons are produced via the
vector-meson-exchange (VME) mechanism ($\rho$ and $\omega$
exchanges). The corresponding diagrams are depicted in
Fig.~\ref{fig:1}. This model, considered as a tool to extract the
radiative decays of scalars to $\rho$ and $\omega$, was proposed
for CLAS $\gamma p$ experiments at high photon-beam energies
$\ega\sim$ several GeV. In the case of $a_0(980)$ and $f_0(980)$
production near threshold ($\ega\sim 1.5 - 1.6$~GeV), one may also
expect sizable contributions from the Born diagrams shown in
Fig.~\ref{fig:2}.

\begin{figure}
\begin{center}
\includegraphics[width=12.0cm, keepaspectratio]{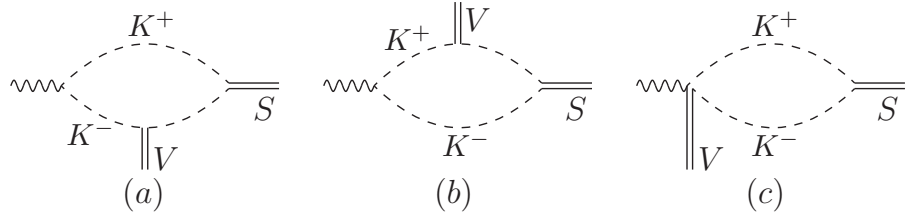} 
\end{center}

\vspace{-4mm}
\caption{Loop diagrams for $SV\gamma$ vertex.
         Dashed lines correspond to charged $K^{\pm}$ mesons.
         Other curves mean the same as in Fig.~\protect\ref{fig:1}.}
	\label{fig:3}
\end{figure}

The diagrams in Fig.~\ref{fig:1} contain an essential ingredient,
i.e., the radiative decay $\ssvg$ vertices ($S=a_0,f_0$; $V=\rho,
\omega$), which can be estimated in different ways. This is the
main source of uncertainties when calculating the diagrams.
Firstly, $SV\gamma$ vertices can be estimated from quark model,
fitted to data on the radiative widths; however, the results
strongly depend on the quark structure of the scalars, which is
not known exactly.  Another approach is the dynamical model for
$SV\gamma$ coupling via intermediate hadronic states. Here,
the main contribution in the case of $a_0(980)$ and $f_0(980)$
comes from the kaon-loop diagrams, shown in Fig.~\ref{fig:3},
which are proportional to the $a_0K\bar K$ or $f_0 K\bar K$
coupling constants. The Born diagrams in Fig.~\ref{fig:2} depend
on the $a_0N\!N$ and $f_0 N\!N$ coupling constants, also known
with large uncertainty.

Further, we consider separately the above-mentioned models and
write down the amplitudes. We use the following notation: $q$,
$p_1$($p_2$), and $k$ are the four-momenta of the initial photon,
initial (final) proton, and final scalar meson $S$, respectively;
$\eps$ is the photon polarization four-vector; $(pq)$ is the scalar
product of four-vectors $p$ and $q$; $\hat p\equiv p_{\mu}\gamu$.

\vspace{3mm}
\centerline{\bf 2.1~Model~A}
\vspace{2mm}

The amplitude $M$ of the reaction $\gamma p\!\to\!pab$ is constructed
from the VME diagrams in Fig.~\ref{fig:1} and reads
\be
	M=\sum_{s=a_0,f_0}M_S,~~~~
	M_S=\sum_{V=\rho,\omega}\frac{1}{t-m^2_V}\,
	\Gamma_{\mu}\,\bar u_2\hat F^{\mu} u_1\,G_S(W)\,g_{sab},
	\label{1}
\ee
where $M_S$ is the amplitude of $S$-meson photoproduction in the
$ab$ channel ($ab=\pi\eta,\,\pi\pi,\,K\bar K$).
Here: $\Gamma_{\mu}$ is the $SV\gamma$ vertex of general (gauge
invariant) form
\be
	\Gamma_{\mu}=I_{\svg}\,[(qk)e_{\mu}-(ek)q_{\mu}],
	\label{2}
\ee
related to the radiative decay width as
\be
	\Gamma(S\to\gamma V)=|I_{\svg}|^2\frac{m^3_S}{32\pi}
	\left(1-\frac{m^2_V}{m^2_S}\right)^3,
	\label{3}
\ee
where $m_V$ ($m_S$) is the mass of vector (scalar) meson;
$F_{\mu}$ is the $VNN$ vertex and
\be
	F_{\mu}=g_V\gamma_{\mu}+if\!_V\sigma_{\mu\nu}p_{\nu}~~ (p=p_2-p_1),
	\label{4}
\ee
where $g_V$ ($f\!_V$) is vector (tensor) $V\!N\!N$ coupling
constant; $u_{1,2}$ are Dirac spinor of the initial and
final nucleons ($\bar uu=2m$, where $m$ is the nucleon mass);
$G_S(W)$, $g_{sab}$ and $W$ are the $S$-meson propagator,
$Sab$-coupling constant and effective mass of meson $ab$
system (the expression for $G_S(W)$ and definition for $g_{sab}$
are given in Appendix~A.1).  The vector-meson propagator in
Eq.~(\ref{1}) is taken as the simple form $1/(t-m^2_V)$, where
$t=(p_2-p_1)^2$, instead of the reggeized prescription used in
Ref.~\cite{Donn}, since we consider the photoproduction of
scalar mesons in the threshold region.$\!\!$
\footnote{~We have already used in Eq.~(\ref{1}) the replacement
$-g_{\mu\nu}+\frac{p_{\mu}p_{\nu}}{m^2_V}\to -g_{\mu\nu}$ for
the numerator of the vector-meson propagator, which is valid
if both nucleons in the VNN vertex are on-shell.}
%
For the $V\!N\!N$ constants ($V\!=\!\rho,\omega$) in Eq.~(\ref{4})
we use the values
\be
	g_{\rho}=3.4,~~ f_{\rho}=11~{\rm GeV}^{-1},~~~
	g_{\omega}=15,~~ f_{\omega}=0.
	\label{41}
\ee
These values were used in Ref.~\cite{Donn} and are consistent
with the description of pion photoproduction~\cite{Laget}.

In the usual definitions the differential \crss\ for $\gamma
p\to (ab)p$ reads
\be
	\frac{d^2\sigma}{dWd\Omega}=\frac{\overline{|M|^2}q_{ab}Q}{(4\pi)^4 Q_{1\,}s},
~~~~~
	\frac{d^2\sigma}{dWdt}=\frac{\overline{|M|^2}q_{ab}}{4(4\pi)^3 Q^2_{1\,}s}.
	\label{5}
\ee
Here, $\overline{|M|^2}$ is the modulus squared of the amplitude
for the unpolarized beam photon and nucleons, and its expression
obtained from Eqs.~(\ref{1})-(\ref{4}) is given in Appendix~A.2;
$d\Omega$ is the solid-angle element of the outgoing $ab$ system in
the reaction rest frame;  $s=m(m+2\ega)$ is the CM total energy
squared, $q_{ab}$ is the relative momentum in the $ab$ system;
$Q_1$ ($Q$) is the momentum of the initial photon (final $ab$ system)
in the reaction rest frame; the additional factor $\half$ is
implied in the case with identical final-state mesons $a$ and $b$. 
The differential \crs\ ~(\ref{5}) is written for the final $ab$
system in the $s$-wave.

In this model, the factor $I_{\svg}$ in Eq.~(\ref{2}) is assumed to
be constant.
We use the width
$\Gamma(S\to\gamma V)$ in Eq.~(\ref{3}) as input to obtain the 
factor $I_{\svg}$.

\vspace{3mm}
\centerline{\bf 2.2~Model~B}
\vspace{2mm}

In this variant, we use the loop mechanism with intermediate hadrons
to calculate the vertex $\Gamma_{\mu}$~(\ref{2}) and factor $I_{\svg}$.
For the scalars $a_0(980)$ and $f_0(980)$, both connected with $K\bar K$
channel, the dominant contribution comes from the $K\bar K$-loop 
diagrams ($a$), ($b$), and ($c$) shown in Fig.~\ref{fig:3}.   The 
diagrams $(a)$ and ($b$) give equal contributions. The third term ($c$), 
which contains the $\gamma V\!KK$ vertex, is prescribed by gauge 
invariance. Also due to this term, the divergencies of the loop 
diagrams in Fig.~\ref{fig:3} are totally cancelled, and one arrives 
at a finite expression for the vertex $\Gamma_{\mu}$~(\ref{2}). Note 
that the result can be obtained, calculating the term, proportional 
to $(ek)q_{\mu}$ in Eq.~(\ref{2}), which comes from diagrams ($a$) 
and ($b$) and is convergent. Calculations were done in
Refs.~\cite{Close,AchaKK} and give
\be
	I_{\svg}=\frac{e g_{SK^+K^-}g_{VK^+K^-}}{2\pi^2 m^2_K}\,I(a,b),~~~~
	a=\frac{m^2_V}{m^2_K},~~~ b=\frac{m^2_S}{m^2_K}.
	\label{6}
\ee
Here, $e$ is the electron charge ($e^2/4\pi\approx 1/137$); $g_{SK^+\!K^-}$
and $g_{V\!K^+\!K^-}$ are the coupling constants of scalar ($S$) and vector 
($V$) mesons to the $K^+K^-$ channel; $m_K$ is the charged kaon mass. The 
function $I(a,b)$ comes from the calculation of the loop integral and is 
given in Appendix~A.3. In the case of VME diagram in Fig.~\ref{fig:1}, the 
vector-meson mass squared $m^2_V$ is replaced by the four-momentum transfer 
$t$, i.e., $a=t/m^2_K$ in Eq.~(\ref{6}).

The constants $g_{SK^+K^-}$ are taken from Refs.~\cite{KLO1,KLO2}, and are
given in Appendix~A.1. For the values of the couplings $g_{VK^+K^-}$, we use
predictions from SU(3) symmetry. Thus,
\be
	g_{\rho K^+K^-}=g_{\omega K^+K^-}=\half\,g_{\rho\pi\pi},~~~~
	\Gamma(\rho\!\to\!\pi\pi)=\frac{g^2_{\rho\pi\pi}q^3_{\pi\pi}}{24\pi m^2_{\rho}}.
	\label{61}
\ee
Here, the constant $g_{\rho\pi\pi}$ is determined in a usual way through
the width and mass of the $\rho$ meson, taken from PDG~\cite{PDG}.

\vspace{3mm}
\centerline{\bf 2.3~Model~C}
\vspace{2mm}

We also estimate the \crss\ from Born diagrams of $a_0/f_0$ photoproduction,
shown in Fig.~\ref{fig:2}.  The amplitude reads
\be
	M=\bar u_2\left[ a_s (\hat p_1\!+\hat q+m)\hat\eps\,
                +\,a_u\hat\eps (\hat p_1\!-\hat k+m)\right]u_1,
	\label{7}
\ee
where
$$
	a_s=\frac{eC}{s-m^2},~~ a_u=\frac{eC}{u-m^2},~~
	C=\sum_{s=a_0,f_0} g_{s\!N}g_{sab}G_S(W),
$$ $$
	s=(p_1+q)^2,~~ u=(p_1-k)^2,
$$
The amplitude is sensitive to the $a_0N\!N$ and $f_0N\!N$ coupling
constants $g_{sN}$ ($s=a_0,f_0$), which are known with large
uncertainty.  For our estimations we take some ``typical"
values~\cite{Faessler}
\be
	g_{a_0N\!N}\simeq g_{f_0N\!N}\simeq 5.
\label{71}\ee
The amplitude squared $\overline{|M|^2}$ for unpolarized photon
and nucleons is given in Appendix~A.2 by Eq.~(A.10).

\vspace{3mm}
\centerline{\bf 2.4~Adding of \bfaf\ mixing}
\vspace{2mm}

\begin{figure}
\begin{center}
\includegraphics[width=8.0cm, keepaspectratio]{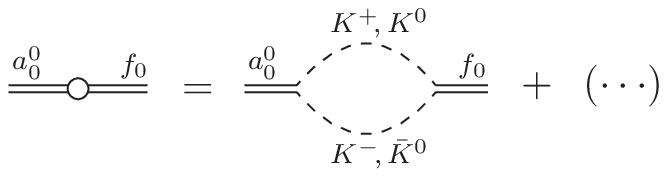}
\end{center}
\vspace{-4mm}
\caption{Diagrammatic representation for $a_0-f_0$ vertex.
         The notation $(\cdots)$ denotes the contributions not
         connected with kaon-loop mechanism and neglected here.}
	\label{fig:4}
\end{figure}

The leading isospin-breaking (IB) effect comes from $a_0-f_0$ mixing. 
Both scalars are coupled to the $K\bar K$ channel, and their masses are
close to the $K\bar K$-threshold. Thus, the contribution to
$a_0\!\leftrightarrow\!f_0$ transition amplitude comes from the mass
difference of charged and neutral kaons and exhibits a sharp maximum
in the 8-MeV mass interval between the $K^+K^-$ and $K^0\bar K^0$
thresholds.  The effect is enhanced since it occurs in the vicinity
of the $a_0$ and $f_0$ masses. The $a_0-f_0$ vertex is shown
diagrammatically in Fig.~\ref{fig:4}. Here, the notation $(\cdots)$
stands for possible terms not connected with the $KK$ loop, assumed
to have a smooth mass dependence. These terms admix contributions
from $f_0(a_0)$ to $a_0(f_0)$ signals, and seem to be not identified
accurately from photoproduction experiments due to the proximity of
the $a_0$ and $f_0$ parameters. On the other hand, the $KK$-loop
term, due to its sharp behaviour, should exhibit a visible signal in
the effective mass spectra in the $a_0$- and $f_0$-decay channels.
The $a_0f_0$ vertex $\lambda$, associated with the $KK$-loop diagram in
Fig.~\ref{fig:4}, reads
\be
	\lambda=i\,\frac{g_{a_0K^+\!K^-}\,g_{f_0K^+\!K^-}}{16\pi m_K}
	\,(q_{K^+K^-}-q_{K^0\bar K^0}),
~~~~
	m_K\!=\half(m_{K^0}\!+m_{K^+}\!),
	\label{8}
\ee
$$
	q_{K^+K^-}=\sqrt{m_K(W\!-2m_{K^+})+i0},~~~~
	q_{K^0\bar K^0}=\sqrt{m_K(W\!-2m_{K^0})+i0}.
$$

The value $|\lambda|$ is maximal at $m_{K^+}\!<\!W\!\!<\!m_{K^0}$, where
$|\lambda|=\frac{g_{a_0K^+\!K^-}\,g_{f_0K^+\!K^-\!}}{8\pi} \sqrt{\frac{m_{K^0}\!
-m_{K^+}}{2m_K}}$, and rapidly decreases beyond this range.

\begin{figure}
\begin{center}
\includegraphics[width=8.0cm, keepaspectratio]{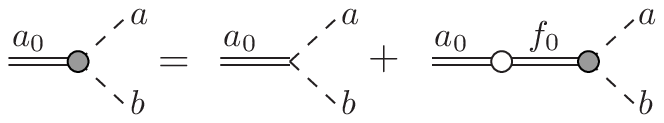}
\end{center}
\vspace{-3mm}
\caption{Diagrammatic representation for rederfined couplings
         $\bar g_{a_0ab}$ and $\bar g_{f_0ab}$ (gray circles),
	including $a_0-f_0$ mixing. The 2-nd equation (not shown)
	mean the replacement $a_0\leftrightarrow f_0$.}\label{fig:5}
\end{figure}

One may include $a_0-f_0$ mixing by replacing the coupling constants $g_{sab}$ of
the scalars to meson $ab$ channels by modified values $\bar g_{sab}$ according
to the relations
\be
	\bar g_{a_0ab}=g_{a_0ab}-\lambda G_f\,\bar g_{f_0ab},~~~~~
	\bar g_{f_0ab}=g_{f_0ab}-\lambda G_a\,\bar g_{a_0ab},
	\label{9}
\ee
which are shown diagrammatically in Fig.~\ref{fig:5}. From Eqs.~(\ref{9}),
we arrive at
\be
	\bar g_{a_0ab}\!=(g_{a_0ab}-\!\lambda G_f\,g_{f_0ab})Z^{-1},~~~
	\bar g_{f_0ab}\!=(g_{f_0ab}-\!\lambda G_a\,g_{a_0ab})Z^{-1},~~~
	Z=1\!-\!\lambda^2 G_a G_f.
	\label{10}
\ee
At $Z=1$, Eqs.~(\ref{10}) include only leading-order terms in the $a_0f_0$
vertex $\lambda$. The redefined vertices $\bar g_{sab}$ as well as the
factors $G_a$, $G_f$ and $\lambda$, depend on the mass $W$, i.e.,
$\bar g_{sab}\equiv\bar g_{sab}(W)$.


\vspace{6mm}
\centerline{\bf 3.~Results}
\vspace{2mm}

Here, we present some results for the total \crss\ and effective $ab$-mass
spectra in different channels $\gamma p\to (ab)p$ , estimated in the models
of Section~2. We calculate only resonance amplitudes, i.e., the final $ab$
system is produced from $a_0/f_0$ decays, and neglect any possible
background terms to the $\gamma p\to (ab)p$ amplitudes. However, it is
interesting to compare the results for the \crss\ obtained using different
approaches.

\vspace{2mm}
\centerline{\bf 3.1~Cross sections}
\vspace{2mm}

First, we obtain the predictions from Model~A; then, we have the radiative
widths $\Gamma(S\to\gamma V)$ to input into the calculation of the $\ssvg$
vertices $I_{\svg}$ in Eq.~(\ref{2}). Here, we may use the results from the
quark model used in Ref.~\cite{Donn}. In this model, assuming the $a_0$
and $f_0$ mesons to be $q\bar q\,(^3P_0)$  states, one obtains results which
depend on the $q\bar q$ flavor configuration.  For the isovector state
$a^0_0=(u\bar u-d\bar d)/\sqrt{2}$~ this gives
\be
	\Gamma(a_0\to\gamma\omega)=125~{\rm keV},~~~
	\Gamma(a_0\to\gamma\rho)=\frac{1}{9}\,\Gamma(a_0\to\gamma\omega)=14~{\rm keV}.
	\label{11}
\ee
Considering three different $q\bar q$ configurations for the isoscalar $f_0$ meson,
denoted as
\be
	f_0(1)=\frac{1}{\sqrt{2}}\,(u\bar u\!+\!d\bar d),~~ 
	f_0(2)=\frac{1}{\sqrt{6}}\,(u\bar u\!+\!d\bar d\!-\!2s\bar s),~~ 
	f_0(3)=\frac{1}{\sqrt{3}}\,(u\bar u\!+\!d\bar d\!+\!s\bar s),
	\label{12}
\ee
one has
\be
	\ba{l}
	\Gamma(f_0(1)\!\to\!\gamma\rho)=3\,\Gamma(f_0(2)\!\to\!\gamma\rho)=
	\frac{3}{2}\Gamma(f_0(3)\!\to\!\gamma\rho)=\Gamma(a_0\!\to\!\gamma\omega),
	\\ \rptt
	\Gamma(f_0(1)\!\to\!\gamma\omega)=3\,\Gamma(f_0(2)\!\to\!\gamma\omega)=
	\frac{3}{2}\Gamma(f_0(3)\!\to\!\gamma\omega)=\Gamma(a_0\!\to\!\gamma\rho).
	\ea
	\label{13}
\ee
The case $f_0=s\bar s$ gives 
$\Gamma(f_0\!\to\!\gamma\rho/\omega)\sim\sin^2\theta$, where $\theta$ is 
the $\phi-\omega$-mixing angle. The angle $\theta$ is assumed to be small 
and this case is not considered here.

The \crss\ $\sigma(ab)$ (in $\mu b$) for different channels are shown in 
Table~1.
The results of Model~A are given for three variants 1), 2), and 3), where
$f_0$ is taken as the $f_0(1)$, $f_0(2)$ and $f_0(3)$ states, respectively.
The \crs\ $\sigma(\pi^0\eta)$ slightly depends on the variants of the $f_0$
states (\ref{12}) due to $a_0-f_0$ mixing, while the main contribution comes
from the $a_0$-production amplitudes. The \crss\ $\sigma(\pi\pi)$ are mainly
determined by the $f_0$-production terms, and are more sensitive to the
$f_0$ variants (\ref{12}).  Models~B and C give comparable values for the
\crs\ but much smaller ones than those obtained from Model~A.

\vspace{2mm}
\begin{center}
\centerline{Table 1}
\vspace{1mm}
%
\centerline{Cross section $\sigma(ab)$ in $\mu b$ for the photoproduction
 of the meson pair $ab$}
\centerline{via $f_0$ and $a_0$ at $\ega=1.6$~GeV, for the models
 described in the text.}
\vspace{2mm}

\begin{tabular}{|c|c|c|c|c|}
\hline\rpt Model & $\pi^0\eta$ &  $\pi^0\pi^0$ & $\pi^+\pi^-$ & $K^+K^-$
\\
\hline\rpt ~~~~1) & 12.21 & 2.59 & 5.12 & 2.58 \\
 A,~2) & 12.08 & 0.86 & 1.70 & 1.28 \\
 ~~~~3) & 12.15 & 1.73 & 3.42 & 1.97
\\
\hline\rpt B & 0.234 & 0.093 & 0.184 & 0.083
\\
\hline\rpt C & 0.444 & 0.070 & 0.138 & 0.072
\\ \hline\end{tabular}\end{center}

\begin{figure}
\begin{center}
\includegraphics[width=12.0cm, keepaspectratio]{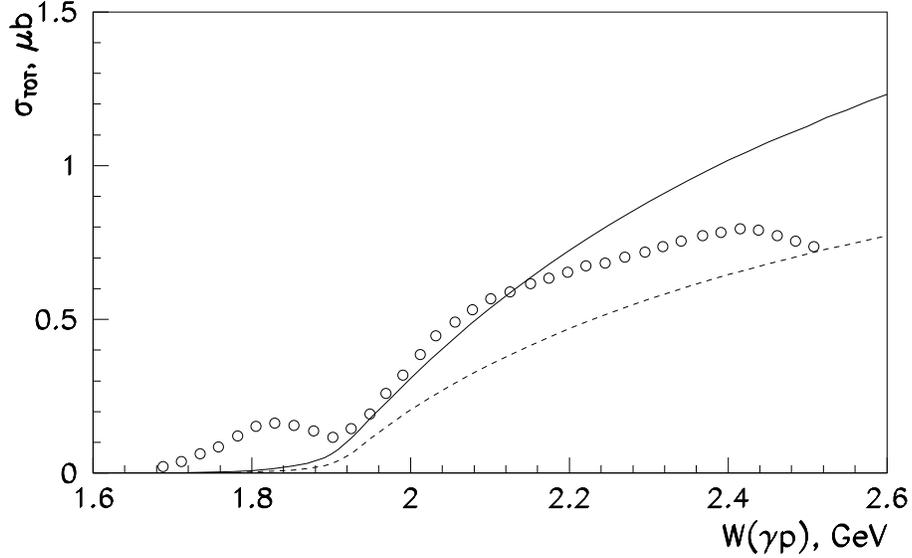}
\end{center}
\caption{Total cross section for $\gamma p\to a^0_0 p\to \pi^0\eta p$
        versus total centre-of-mass energy $W(\gamma p)$.
        The curves show the results from the Model~B
        ($a_0-f_0$ mixing is included). 
	The results are given for two sets of $a_0/f_0$ parameters,
	taken from Refs.~\protect\cite{KLO1,KLO2} -- ``kaon loop"
	fit (solid curve) and ``no structure" fit (dashed curve)
	(see Eq.~(A.4)).
	Open circles show the $a_0 p$ contribution to the $\gamma
	p\to \pi^0\eta p$ cross section extracted through PWA in
	Ref.~\protect\cite{Horn}.
}\label{fig:as}
\end{figure}

For some comparison with existing data we present in Fig.~\ref{fig:as}
the cross section $\sigma(\pi^0\eta)$ versus total centre-of-mass energy
$\sqrt{s}=W(\gamma p)$ in Model~B for two sets of $a_0/f_0$ parameters
(solid and dashed curves, see figure caption).  Here, open circles show
the $a_0(980)p$ contribution to the $\pi^0\eta p$  channel, obtained
through a partial-wave analysis (PWA) of the data on $\gamma p\to\pi^0\eta
p$ in Ref.~\cite{Horn}. Thus, the model results essentially depend on
$a_0/f_0$ parameters, but Model~B is in rough agreement with the ``data"
(open circles). The cross sections from Model~A are too large and not
shown in Fig.~\ref{fig:as}. Note also that $W(\gamma p)=1.97$~GeV for
the energy of interest $\ega=1.6$~GeV.

Concerning the other channels, there are recent CLAS high-statistics data
on the reaction $\gamma p\to \pi^+\pi^- p$ at $\ega=3.0-3.8$~GeV~\cite{Batta}. 
The PWA results of Ref.~\cite{Batta} give, in particular, the contribution
of the $s$-wave system $(\pi^+\pi^-)_s$ with clear evidence of the $f_0(980)$
structure.  One should also mention the old hydrogen $\gamma p\to K^+K^-p$
data~\cite{Barber}, where the $s$-wave $(K^+K^-)_s$ cross section and
possible contributions of a scalar resonance ($M_{\pi\pi}\sim 1$~GeV) were
estimated.

Generally, the photoproduction processes of scalars should be analyzed in
the full approach which incorporates the resonance and background terms
and utilizes unitarity. For example, the background tree
$\rho,\omega$-exchange amplitudes for $\pi\pi$, $\pi\eta$, and $\eta\eta$
channels were taken into account in Ref.~\cite{Donn}, and their contribution
was found to be comparable with the resonance terms in the corresponding mass
intervals. Analogous tree amplitudes supplemented with $s$-wave meson-meson
final state interaction (FSI) were considered for the $\pi\pi$ and $K\bar K$
photoproduction in Refs.~\cite{Szczepan}, where the cross sections for
$s$-wave $(\pi\pi)_s$ and $(K\bar K)_s$ pairs were estimated.

In the present paper, we leave the inclusion of such background
processes for future work and study the $a_0-f_0$-mixing effect which
is produced by the resonance amplitudes.

\vspace{3mm}
\centerline{\bf 3.2~Mass spectra}
\vspace{2mm}

The results for the total \crss\ given in Table~1 exhibit a strong
model dependence, but are only weakly sensitive to the $a_0-f_0$ mixing.  As
mentioned above, the $a_0f_0$ vertex $\lambda$~(\ref{8}) sharply depends
on the mass $W$ and peaks close to the $K\bar K$ thresholds.

\begin{figure}
\begin{center}
\includegraphics[width=15.0cm, keepaspectratio]{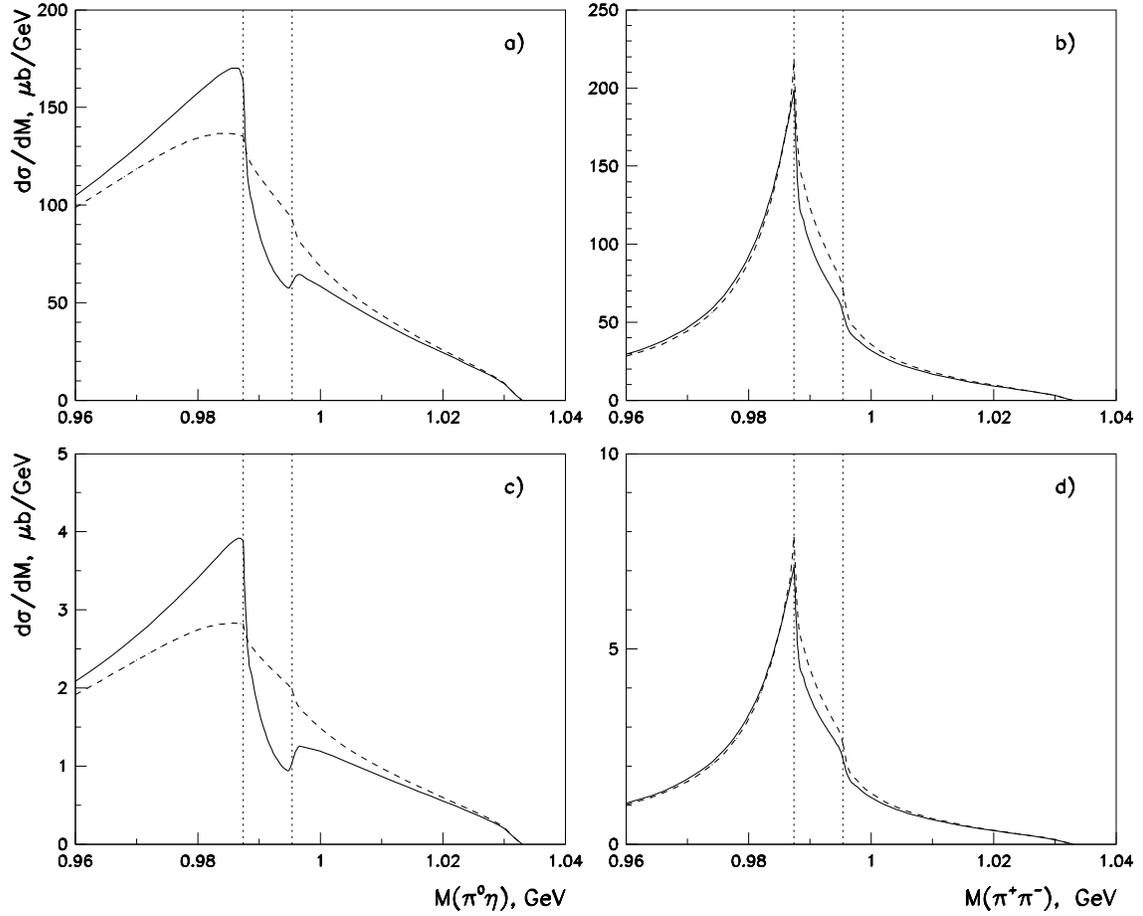}
\end{center}
\caption{The mass distributions $d\sigma/dM(\pi^0\eta)$ (plots $a$ and $c$)
        and $d\sigma/dM(\pi^+\pi^-)$ (plots $b$ and $d$) in the reactions
        $\gamma p\to(\pi^0\eta)p$ and $\gamma p\to(\pi^+\pi^-)p$,
	respectively, at $\ega=1.6$~GeV. The plots $a$ and $b$ show the
	results from the Model~A with variant $f_0=f_0(1)$~(\ref{12});
	the plots $c$ and $d$ -- the results from the Model~B. Solid
	(dashed) curves show the results obtained with $a_0-f_0$ mixing
	included (excluded).  The $a_0/f_0$ parameters are taken from
	Refs.~\protect\cite{KLO1,KLO2} (``kaon loop" fits).
        Vertical dotted lines point the $K\bar K$-threshold positions.
	}\label{fig:6}
\end{figure}

The mass spectra for the two channels $\pi^0\eta$ and $\pi^+\pi^-$ at the
beam-photon energy $\ega=1.6$~GeV are presented in Fig.~\ref{fig:6}. The
results are given for two models, A (variant 1) and B with the $a_0/f_0$
parameters from the ``KK" version~(see, Eqs.~(A.4)).  All the plots in
Fig.~\ref{fig:6} exhibit two kinds of phenomena: the ``cusp" effects at
the $K\bar K$ thresholds and $a_0-f_0$-mixing. The latter is seen as the
differences of solid and dashed curves. The ``cusp" effects look more
pronounced in the $\pi\pi$ channel than in the $\pi\eta$ one, essentially
because the $f_0$ has larger coupling to the $K\bar K$ channels than the 
$a_0$.

\begin{figure}
\begin{center}
\includegraphics[width=15.0cm, keepaspectratio]{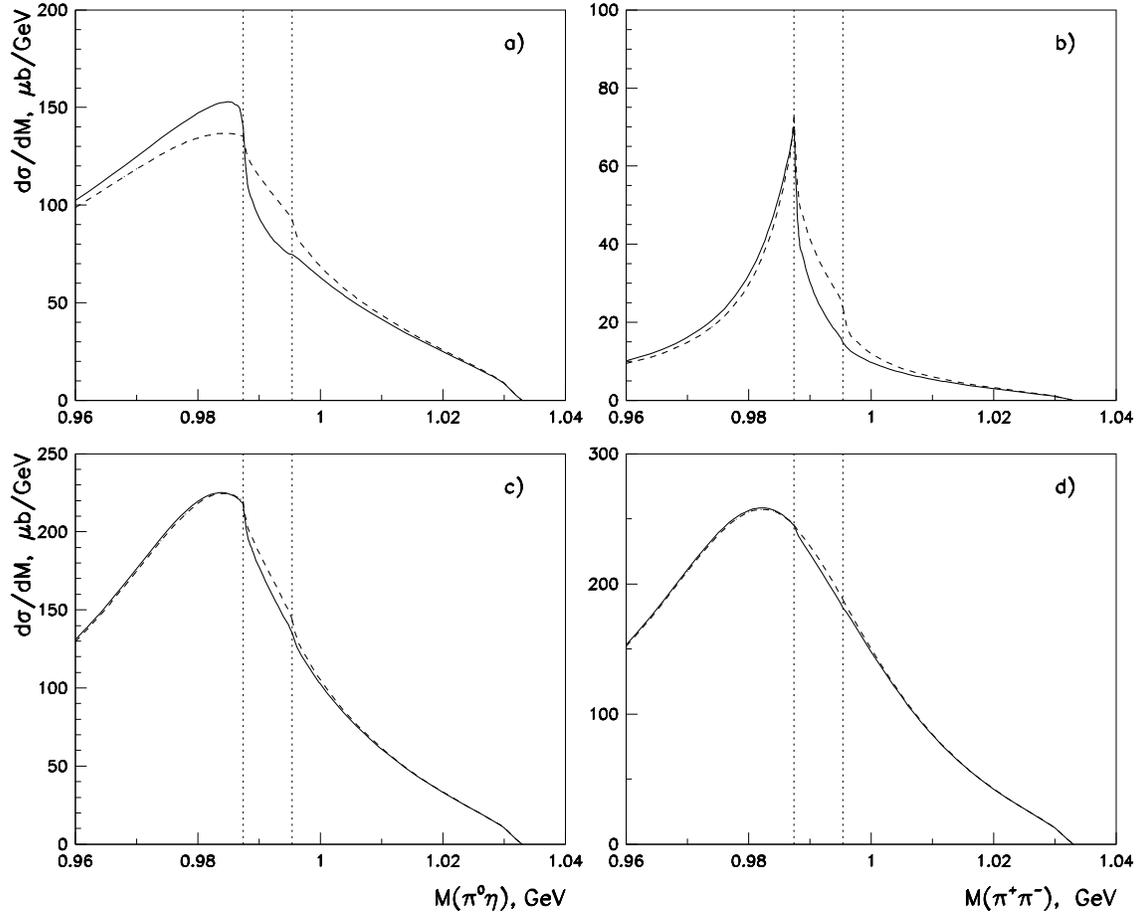}
\end{center}
\caption{The mass distributions $d\sigma/dM(\pi^0\eta)$ (plots $a$ and
	$c$) and $d\sigma/dM(\pi^+\pi^-)$ (plots $b$ and $d$),
	respectively.  The reactions are the same as in Fig.~\ref{fig:6},
	and $\ega=1.6$~GeV.
        The plots $a$ and $b$ show the results from the Model~A with
	variant $f_0=f_0(2)$~(\protect\ref{12}) with the $a_0/f_0$
	parameters from Refs.~\protect\cite{KLO1,KLO2} (``kaon loop"
	fits); the plots $c$ and $d$ -- the results from the Model~A
	with variant $f_0=f_0(1)$~(\protect\ref{12}) with the $a_0/f_0$
	parameters from Refs.~\protect\cite{KLO1,KLO2} (``NS" fits).
	Notations of the curves are the same as in
	Fig.~\protect\ref{fig:6}.}\label{fig:7}
\end{figure}

Models~A and B in Fig.~\ref{fig:6} give quite similar shapes of mass
spectra. To get some view of model dependence of the results, we present
some other predictions for the same channels in Fig.~\ref{fig:7}.  The
plots $a$ and $b$ show the mass spectra obtained in Model~A (variant 2,
i.e., $f_0=f_0(2)$ in Eq.~(\ref{12})) with the same ``KK" variant of
the $a_0/f_0$ parameters. Here, since the radiative widths
$\Gamma(f_0\!\to\!\gamma\rho/\omega)$ for $f_0(2)$ is 3 times smaller
than for $f_0(1)$, the $d\sigma/dM(\pi\pi)$ is also getting $\sim 3$
smaller in comparison with that in plot $b$ of Fig.~\ref{fig:6}.

The plots $c$ and $d$ in Fig.~\ref{fig:7} show the results from Model~A
(variant~1), but with the no-structure (``NS") variant of the $a_0/f_0$ 
parameters. In the ``NS" version, both constants $g_{aK^+K^-}$ and 
$g_{fK^+K^-}$ are smaller (the latter by $\sim$ one order of magnitude) 
than their ``KK"- version values (see Eqs.~(A.4)). Thus, the ``cusp" 
effects as well as the $a_0-f_0$-mixing (note that the $a_0f_0$ 
vertex~(\ref{8}) $\lambda\sim g_{aK^+K^-}g_{fK^+K^-}$) are hardly 
visible in this case.

\begin{figure}
\begin{center}
\includegraphics[width=13.0cm, keepaspectratio]{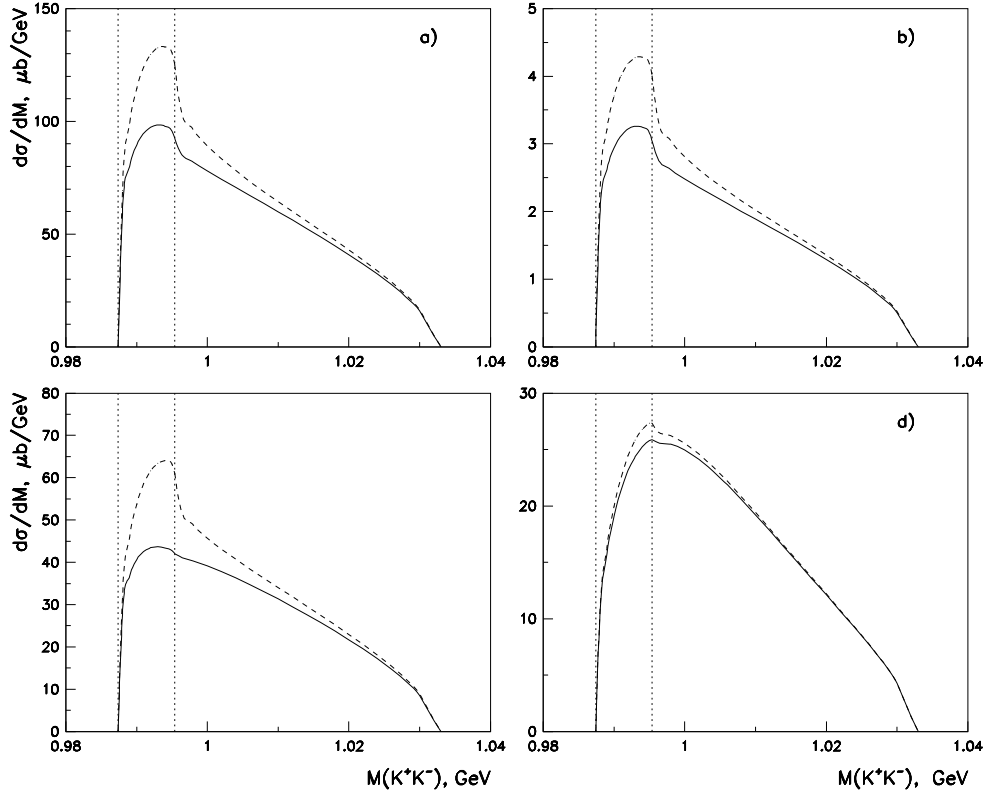}
\end{center}
\caption{The mass distributions $d\sigma/dM(K^+K^-)$ in the reaction
        $\gamma p\to(K^+K^-)p$ at $\ega=1.6$~GeV.
	The plots: $a,d$ -- model A with variant
	$f_0=f_0(1)$~(\protect\ref{12});
	$b$ -- Model~B; $c$ -- Model~A with variant
	$f_0=f_0(2)$~(\protect\ref{12}).
	The $a_0/f_0$ parameters are taken from
	Refs.~\protect\cite{KLO1,KLO2}:
	plots $a,b,c$ -- ``kaon loop" fits; $d$ -- ``NS" fits.
	Notations of the curves are the same as in
	Fig.~\protect\ref{fig:6}.}\label{fig:8}
\end{figure}

Fig.~\ref{fig:8} shows the effective $K^+K^-$ mass spectra in the reaction
$\gamma p\!\to\!(K^+K^-)p$ at the same photon energy. Here we give the
results of the same four variants of the model calculations as in
Figs.~\ref{fig:6} and~\ref{fig:7} (see figure caption).  Plot $d$ shows
the results obtained with the ``NS" version of the $a_0/f_0$ parameters.
$a_0-f_0$-mixing is also suppressed here due to smaller couplings of
the resonances to the $K\bar K$ channels.  Thus, we see that the IB
$a_0-f_0$-mixing effect essentially depends on the $a_0/f_0$
parameters, in particular on the $a_0$ and $f_0$ couplings to the
$K\bar K$ channel.

  From an experimental point of view, one can not measure the reaction
discussed with ``switched off" isospin-breaking effects in order to
observe any difference in the mass spectra like those between the
solid and dotted curves in Figs.~\ref{fig:6}-\ref{fig:8}. Thus, we
also need to study the charged channels, where mixing is absent, say,
$a^+_0$ photoproduction in $\gamma p\!\to\!(\pi^+\eta)n$, in parallel
with the neutral channels to compare the results.

\vspace{5mm}
\centerline{\bf 4.~Conclusion}
\vspace{3mm}

The photoproduction of the neutral scalars $a_0(980)$ and $f_0(980)$ on
a proton target at energies close to threshold were considered in
the $\pi\eta$, $\pi\pi$, and $K\bar K$ channels. The main aim of the
paper is to study the possibility of observing $a_0-f_0$ mixing in
these processes.  Several models of $a_0/f_0$ photoproduction were
considered with $a_0-f_0$ mixing included through the $K\bar K$-loop
mechanism of $a_0-f_0$ transition. The total cross sections of
different channels were estimated and appeared to be very model
dependent. Model~B, incorporating $\rho$ and $\omega$-exchange
diagrams and a $K\bar K$-loop mechanism for $a_0/f_0$ photoproduction,
demonstrates rough agreement with the data on the $a_0$ contribution
to the $\gamma p\to\pi^0\eta p$ cross section.

The two-meson mass spectra are examined for observation of
$a_0-f_0$-mixing. The most interesting case is the $\gamma p\to\pi^0\eta
p$ channel.  Here, the $\pi^0\eta$-effective-mass spectrum demonstrates
a sharp (mixing) effect (Fig.~\ref{fig:6}), i.e., rapid behavior of
the $d\sigma/dM$ in the narrow ($\sim 8$~MeV) mass interval, for the
case of $a_0/f_0$ parameters, taken from ``kaon loop" fits of
Refs.~\cite{KLO1,KLO2}. The effect is sensitive to the $a_0/f_0$
parameters.

Both aspects, the $\ssvg$ vertex $I_{\svg}$ (see Eq.~(\ref{2})), which
affects the photoproduction cross section of scalars, and the
$a_0-f_0$-mixing vertex $\lambda$ (\ref{8}), are important to understand
the nature of scalar mesons $a_0(980)$ and $f_0(980)$. We expect this
study to be continued in a more complete model, incorporating also
the background amplitudes for the given channels. 

\vspace{3mm}
\centerline{\bf Acknowledgments}
\vspace{3mm}

The authors are thankful to M.~Amaryan and S.~Prakhov for many
useful discussions concerning the experimental possibilities to
measure $a_0/f_0$ photoproduction and E.~Oset for useful references. 
This work was supported in part by the U.S. Department of Energy 
Grant No.~DE--FG02--99ER41110 and the DFG under grant SFB 1044. 
A.~E.~K. thanks grant NS--3172.2012.2 for partial support.

\vspace{3mm}
\begin{flushright}
{\bf Appendix}
\end{flushright}

\centerline{\bf A.1~Scalar meson propagators}
\vspace{2mm}

The propagators of scalars reads
$$
	G_S=(W^2-m^2_S+iW\,\Gamma_S(W))^{-1},
~~~~
	\Gamma_S(W)=\sum_{ab}\Gamma_{sab}(W).
\eqno{(\rm A.1)}
$$
The total width $\Gamma_S(W)$ is the sum of partial widths
$\Gamma_{sab}(W)$ of the $s$-wave decays $S\!\to\!ab$, and
$$\Gamma_{\!sab}(W)=\frac{g^2_{sab} q_{ab}}{8\pi W^2},~~~~~
ab=\biggl\{
\ba{ll}
	\pi^0\eta,\,K^+\!K^-\!,\,K^0\bar K^0 & (S=a^0_0) \\
	\pi\pi\!,~~\,K^+\!K^-\!,\,K^0\bar K^0 & (S=f_0)
\ea.
	\eqno{(\rm A.2)}
$$
Here, $g_{sab}$ is the coupling constant of the scalar $S$ to $ab$ channel;
$$
	q_{ab}=\sqrt{\frac{1}{4W^2}\,(W^2\!-m^2_+)(W^2\!-m^2_-)+i0},~~~~~
	m_{\pm}=m_a\pm m_b,
\eqno{(\rm A.3)}
$$
where $q_{ab}$ is the relative momentum in the $ab$ system with
effective mass $W$, and $m_a$ ($m_b$) is the mass of particle $a$ ($b$).
The value $q_{ab}$ in Eq.~(A.3) is also defined in the region below
threshold, i.e., $q_{ab}=i|\,q_{ab}|$ at $W\!<\!m_a\!+\!m_b$.

The $a_0(980)$ and $f_0(980)$ parameters are taken from the analyses
of $\phi(1020)\!\to\!\pi^0\eta\gamma$~\cite{KLO1} and
$\phi\!\to\!\pi^0\pi^0\gamma$~\cite{KLO2}. The results were obtained
for two variants of fits -- ``kaon loop" (``KK") and ``no structure"
(``NS") models:

$$
\ba{llcl}
{\rm ``KK"}\!\!: & m_a\!=983~{\rm MeV}, & g_{a\pi\eta}=2.8~{\rm GeV}, &
                     g_{aK^+K^-}\!=2.16~{\rm GeV};
\\ \rptt
{\rm ``KK"\!\!}: & m_f\!=976.8~{\rm MeV}, & g_{f\pi^+\pi^-}\!=-1.43~{\rm GeV}, &
                  g_{fK^+K^-}\!=3.76~{\rm GeV};
\\ \rptt
{\rm ``NS":} & m_a\!=983~{\rm MeV}, & g_{a\pi\eta}=2.2~{\rm GeV}, &
                     g_{aK^+K^-}\!=1.57~{\rm GeV};
\\ \rptt
{\rm ``NS"}\!: & m_f\!=984.7~{\rm MeV}, & g_{f\pi^+\pi^-}\!=1.31~{\rm GeV}, &
g_{fK^+K^-}\!=0.40~{\rm GeV}
\ea
\eqno{(\rm A.4)}
$$
($g_{aK^0\bar K^0}\!=-g_{aK^+K^-}$,~ $g_{fK^0\bar K^0}\!=g_{fK^+K^-}$).

\vspace{3mm}
\centerline{\bf A.2~Reaction amplitude squared}
\vspace{2mm}

{\bf Models~A,B}.
The reaction amplitude $M$ can be written as
$$
	M=\bar u_2(A+\hat B)u_1,
	\eqno{(\rm A.5)}
$$
where
$$
	A=a_1[(\eps k)(q,p_1+p_2)-(qk)(\eps,p_1+p_2)],~~~~
	\hat B=a_3[(\eps k)\hat q-(qk)\hat\eps],
$$ $$
	a_1=\sum_{S,V} \frac{G_S\, g_{sab}f_V}{t-m^2_V}\,I_{SV},~~~
%
	a_2=\sum_{S,V} \frac{G_S\, g_{sab}g_V}{t-m^2_V}\,I_{SV},~~~
	a_3=2ma_1+a_2.
$$
The modulus squared of the amplitude for unpolarized particles
reads
$$
	\overline{|M|^2}=\frac{1}{2}\,{\rm Tr}\,
	\{(A^*\!+\hat B^*)(\hat p_2\!+m)((A\!+\hat B)(\hat p_1\!+m)\}
	~~~ (\hat B^*\!\equiv B^*_{\mu}\gamu),
	\eqno{(\rm A.6)}
$$
where the trace Tr$\{\cdots\}$ is averaged over photon polarizations.
To simplify calculations we impose gauge condition $\eps_0\!=\eps_3\!=0$
on the photon four-vector $\eps$. Thus, the total set of useful scalar
products with four-vector $\eps$ is
$$
	(\eps q)=(\eps p_1)=0,~~ (\eps p_2)=-(\eps k)=(\beps \bfk_{\perp}),~~ \eps^2=-1.
	\eqno{(\rm A.7)}
$$
Finally, from Eq.~(A.6), making use of Eq.~(A.7), we arrive at
$$
	\overline{|M|^2}=2(\,|\,a_2|^2\!-|\,a_1|^2 t) (qp_{1\!})^2 k^2_{\perp}-|\,a_3|^2 (qk)^2 t,
	\eqno{(\rm A.8)}
$$
where substitution $(\eps k)^2\to\half k^2_{\perp}$ is used for unpolarized photon.
The factors $(qp_{1\!})^2 k^2_{\perp}$ and $(qk)^2$ in Eq.~(A.8) can be written as
$$
(qp_{1\!})^2 k^2_{\perp}=\frac{1}{4}s(t_2-|t|)(|t|-t_1),~~~~
(qk)^2=\half(W^2-t),
$$
where $t_1$ and $t_2$ are the kinematical boundaries for $|t|$ ($t_1\!<\!|t|\!<\!t_2$).

\vspace{2mm}
{\bf Model~C}.
Making use of Eqs.~(A.7) and Dirac equations for nucleon spinors $u_1$ and
$\bar u_2$, one can rewrite the amplitude in Eq.~(\ref{7}) in the form
$$
	M=\bar u_2\left[\,2a_u (\eps p_2)+(a_s+a_u)\,\hat q\hat\eps\,\right]u_1.
	\eqno{(\rm A.9)}
$$
Calculations for unpolarized particles give
$$
	\overline{|M|^2}=4\left[\,|\,a_s\!+\!a_u|^2(qp_1)(qp_2)+
	k^2_{\perp} [\,|\,a_u|^2(m^2\!+(p_1,p_2\!-q))-{\rm Re}(a^+_sa_u)(qp_1)]\right].
	\eqno{(\rm A.10)}
$$


\newpage
\vspace{3mm}
\centerline{\bf A.3~Loop function $I(a,b)$}
\vspace{2mm}

The loop function $I(a,b)$, which enters the $\ssvg$ vertex $I_{svg}$
in Eq.~(\ref{6}), can be written as the integral
$$
	I(a,b)=\int\limits^1_0 dz \int\limits^{1-z}_0 dy \frac{yz}{c-i0},~~~
	c=1-z(1-z)a-yz(b-a).
$$
Calculations give (see, also \cite{Close,Donn})
$$
	I(a,b)=\frac{1}{2(a-b)}+\frac{a}{2(a-b)^2}\,[J(b)-J(a)]
      +\frac{1}{2(a-b)^2}\,[f(b)-f(a)],
	\eqno{(\rm A.11)}
$$
where
$$
\ba{lll}
	1)~J(a)=x(L-i\pi), & f(a)=-(L-i\pi)^2, & (a>4);
	\\ \rptt
	2)~J(a)=2xA, & f(a)=4A^2,  & (0<a<4);
\\ \rptt
	3)~J(a)=xL, & f(a)=-L^2, & (a<0);
	\ea
$$
$$
	x=\sqrt{\left|\frac{a-4}{a}\right|},~~~ L=\ln\left|\frac{1+x}{1-x}\right|,
	~~~A=\arcsin\frac{\sqrt{a}}{2}.
$$


\newpage



\newpage

\end{document}